\documentclass[12pt]{revtex4}
\def\comment#1{}

\usepackage{graphicx}
\begin{document}
\title{The phase structure of Einstein-Cartan theory}
\author{She-Sheng Xue}
\affiliation{ICRANeT Piazzale della Repubblica, 10 -65122, Pescara, Italy}

\begin{abstract}
In the Einstein-Cartan theory of torsion-free gravity coupling to massless fermions, the 
four-fermion interaction is induced and its strength is a function of the gravitational and gauge couplings,
as well as the Immirzi parameter. We study
the dynamics of the four-fermion interaction to determine whether  
effective bilinear terms of massive fermion fields are generated. 
Calculating one-particle-irreducible two-point functions of 
fermion fields, we identify three different phases and two critical points for phase transitions 
characterized by the strength of four-fermion interaction: 
(1) chiral symmetric phase for massive fermions in strong coupling regime; (2) chiral symmetric broken phase 
for massive fermions in intermediate coupling regime; 
(3) chiral symmetric phase for massless fermions in weak coupling regime. 
We discuss the scaling-invariant region for an effective theory of 
massive fermions coupled to torsion-free gravity  
in the {\it low-energy limit}.     

\end{abstract}
\pacs{95.30.Sf,12.60.Cn,12.60.Fr}
\maketitle

\vskip0.5cm
\noindent
{\bf Introduction.}
\hskip0.2cm
The self-dual connection of a Yang-Mills gauge theory introduced in the Ashtekar 
formalism for General Relativity \cite{A1986} is crucial for the canonical quantization procedure, 
leading to the non-perturbative quantum theory of gravity, 
{\it Loop Quantum Gravity} \cite{loopr,loopa,loopt}. The complex Ashtekar's connection with 
reality condition and the real Barbero real connection \cite{b1995} are linked by a 
canonical transformation of the connection with the 
Immirzi parameter $\gamma\not=0$ \cite{i1997},
which has crucial effects on quantum gravity \cite{rt1998} at the Planck energy scale, but does not affect the 
classical dynamics of torsion-free gravity. However, when fermion fields are present 
and coupled to gravity, yielding a non-vanishing torsion tensor \cite{s2001}, and the 
Einstein-Cartan theory for torsion-free gravity coupling to fermions should be modified.
Indeed, Refs.~\cite{ar05} show that the four-fermion interacting strength in the Einstein-Cartan 
theory is related to the Immirzi parameter, which can possibly lead to physical effects observable.
Thus, it is worthwhile to study the dynamics of these quadralinear terms of fermion fields in terms of the four-fermion 
interacting strength to see whether effective bilinear terms of massive fermions are generated.  
  
\vskip0.5cm
\noindent
{\bf Einstein-Cartan theory.}
\hskip0.2cm
\comment{
To derive the Einstein-Cartan theory we use the following notations. ${\mathcal M}$ the 4-dimensional Euclidean space-time manifold, 
In the Ashtekar formalism for General Relativity 
and $g_{\mu\nu}$ space-time matrix with signature $(+,+,+,+)$.
For the tetrad formalism, we fix a four-dimensional vector space $V$ equipped with a fixed metric 
$\eta^{ab}$ of signature $(+,+,+,+)$, which will serve as the `internal space'. Orthonormal co-tetrats will be denoted by
by $e^a_\mu$; thus $g_{\mu\nu}=\eta_{ab}e^a_\mu e^b_\nu$. In this vector space, the conventions of Dirac $\gamma$-matrices are: 
communication $\{\gamma_a,\gamma_b\}=-2\eta_{ab}$; anti-hermitian $\gamma^\dagger_a=-\gamma_a$ and $\gamma^2_a=-1$ ($a=0,1,2,3$). 
The hermitian $\gamma_5$-matrix $\gamma_5^\dagger=\gamma_5$, $\gamma_5=\gamma^5=\gamma^0\gamma^1\gamma^2\gamma^3=\gamma_0\gamma_1\gamma_2\gamma_3$ and $\gamma_5^2=1$.  The hermitian spinor matrix $\sigma^{ab}=\frac{i}{2}[\gamma^a,\gamma^b]$, 
and $\epsilon_{\mu\nu\rho\sigma}=\epsilon_{abcd}e^a_\mu e^b_\nu e^c_\rho e^d_\sigma$ is totally antisymmetric tensor.
}
In the Palatini framework, the basic gravitational variables constitute a pair of tetrat and spin-connection fields 
$(e_\mu^a, \omega^{ab}_\mu)$. They are 1-form fields 
on ${\mathcal M}$ the 4-dimensional Euclidean space-time manifold, taking values, respectively, in the vector 
space $V$ and in the Lie algebra $so(\eta)$ of the group $SO(\eta)$ of the linear transformations of
$V$ preserving $\eta^{ab}=(+,+,+,+)$. The 2-form curvature associating with the spin-connection is
\begin{equation}
R^{ab}=d\omega^{ab} -\omega^{ae}\wedge\omega^{b}{}_{e}.
\label{rcurvature}
\end{equation}
The Palatini action for gravitational field is given by,
\begin{equation}
S_P(e,\omega)=\frac{1}{4k}\int_{\mathcal M} d^4x\det(e)\epsilon_{abcd}e^a\wedge e^b \wedge R^{cd},
\label{host}
\end{equation}
where $k\equiv 8\pi G$. The relationship between spin-connection $\omega^{ab}_\mu$ and the tetrat $e_\mu^a$ is determined by 
$\delta S_P(e,\omega)/\delta \omega=0$, Cartan's structure equation,   
\begin{equation}
de^a -\omega^{ab}\wedge e_{b}=0,
\label{werelation}
\end{equation}
which gives the torsion-free spin-connection: $\omega=\omega(e)$. Replacing $\omega$ in Eq.~(\ref{host}) by 
$\omega=\omega(e)$, the Palatini action $S_P[e,\omega(e)]$ reduces to the Einstein-Hilbert action and 
its variation with respect to the tetrad field $e_\mu^a$ leads to the Einstein field equation,
\begin{equation}
\epsilon_{abcd}e^b\wedge R^{cd}[\omega(e)]=0.
\label{eineq}
\end{equation}
Adding the Host modification with the Immirzi parameter $\gamma$, one has
\begin{equation}
S_H(e,\omega)=S_P(e,\omega)-\frac{1}{2k\gamma}\int_{\mathcal M} d^4x\det(e) e_a\wedge e_b \wedge R^{ab}.
\label{host1}
\end{equation}   
Introducing massless Dirac fermions $\psi$ coupled to the gravitational field described by $(e_\mu^a, \omega^{ab}_\mu)$, 
we adopt the fermion action of Ashtekar-Romano-Tate type \cite{art1989},
\begin{eqnarray}
S_F(e,\omega,\psi,\bar\psi)=\frac{1}{2}\int_{\mathcal M} d^4x\det(e)\left[\bar\psi e^\mu {\mathcal D}_\mu\psi +{\rm h.c.}\right],
\label{art}
\end{eqnarray}
where the covariant derivative 
\begin{equation}
{\mathcal D}_\mu=\partial_\mu - \frac{i}{4}\beta\omega_\mu,
\label{cd}
\end{equation}
$\beta$ is the gauge coupling between fermion and spin-connection fields, 
Dirac-matrix valued tetrad and spin-connection fields are $e^\mu\equiv e^\mu_a\gamma^a$ 
and $\omega_\mu\equiv \omega^{ab}_\mu\sigma_{ab}$. The anti-hermitian Dirac matrix 
$\gamma^\dagger_a=-\gamma_a$, $\gamma^2_a=-1$ ($a=0,1,2,3$),
$\{\gamma_a,\gamma_b\}=-2\eta_{ab}$, and the hermitian 
spinor matrix $\sigma^{ab}=\frac{i}{2}[\gamma^a,\gamma^b]$. 
The actions (\ref{host},\ref{art}) are invariant under the 
diffeomorphisms of the manifold ${\mathcal M}$, and can be separated into 
left- and right-handed parts \cite{lrseparation}, 
with respect to local $SU_L(2)-$ and $SU_R(2)-$ Lorentz symmetries. 
\comment{
This can be shown by writing Dirac fermion 
$\psi=\psi_L+\psi_R$, where Weyl fermions $\psi_{L,R}=P_{L,R}\psi, P_{L,R}=(1\mp\gamma_5)/2$; and 
Dirac-matrix valued tetrad and spin-connection fields $e^\mu=P_Le^\mu+P_Re^\mu$ and $\omega_\mu=P_L\omega_\mu+P_R\omega_\mu$.
}

The Palatini action (\ref{host}) and fermion action (\ref{art}) give the Einstein-Cartan action,
\begin{equation}
S_{EC}=S_P(e,\omega)+S_F(e,\omega,\psi,\bar\psi).
\label{ec}
\end{equation}
Analogously to 
Eq.~(\ref{werelation}), $\delta S_{EC}(e,\omega)/\delta \omega=0$ gives Cartan's structure equation,
\begin{equation}
de^a -\omega^{ab}\wedge e_{b}-T^a=0,
\label{werelation1}
\end{equation}
where the non-vanishing torsion field $T^a=k\beta e_b\wedge e_cJ^{ab,c}$,
relating to the fermion spin-current
\begin{equation}
J^{ab,c}=\frac{i}{4}\bar\psi\{\sigma^{ab},\gamma^c\}\psi=\frac{1}{4}\epsilon^{abcd}\bar\psi\gamma_d\gamma^5\psi,
\label{spinc}
\end{equation}
and $\{\sigma^{ab},\gamma^c\}=i\epsilon^{abcd}\gamma^5\gamma_d$. The hermitian 
$\gamma_5$-matrix 
$\gamma_5=\gamma^5=\gamma^0\gamma^1\gamma^2\gamma^3=\gamma_0\gamma_1\gamma_2\gamma_3$, $\gamma_5^\dagger=\gamma_5$ and $\gamma_5^2=1$. 
The totally antisymmetric tensor $\epsilon_{\mu\nu\rho\sigma}=\epsilon_{abcd}e^a_\mu e^b_\nu e^c_\rho e^d_\sigma$.
The solution to Eq.~(\ref{werelation1}) is 
\begin{equation}
\omega^{ab}_\mu=\omega^{ab}_\mu(e)+\tilde\omega^{ab}_\mu,\quad
\tilde\omega^{ab}_\mu=k\beta e_\mu^cJ^{ab}{}_c,
\label{connectiont}
\end{equation}
where the connection $\omega^{ab}_\mu(e)$ obeys Eq.~(\ref{werelation}) for torsion-free case. The fermion spin-current (\ref{spinc})
contributes only to the pseudo-trace axial vector of torsion tensor, which is one of irreducible parts of torsion 
tensor \cite{torsion2001}.
Replacing the spin-connection $\omega$ in the Einstein-Cartan action (\ref{ec}) by (\ref{connectiont}),
\begin{eqnarray}
S_P[e,\omega,\psi,\bar\psi]&\rightarrow & S_P[e,\omega(e)]-\frac{1}{16}k\beta^2\int_{\mathcal M} d^4x\det(e)(\bar\psi\gamma^d\gamma^5\psi)(\bar\psi\gamma_d\gamma^5\psi);\label{eca0}\\
S_F[e,\omega,\psi,\bar\psi]&\rightarrow & S_F[e,\omega(e),\psi,\bar\psi]
-\frac{2}{16}k\beta^2\int_{\mathcal M} d^4x\det(e)(\bar\psi\gamma^d\gamma^5\psi)(\bar\psi\gamma_d\gamma^5\psi),
\label{eca1}
\end{eqnarray} 
one obtains the well-known Einstein-Cartan theory: 
the standard tetrad action of torsion-free gravity coupling to fermions,
\begin{eqnarray}
S_{EC}[e,\omega(e),\psi,\bar\psi]&=&S_P[e,\omega(e)]+S_F[e,\omega(e),\psi,\bar\psi]\nonumber\\
&-&\frac{3}{16}k\beta^2\int_{\mathcal M} d^4x\det(e)(\bar\psi\gamma^d\gamma^5\psi)(\bar\psi\gamma_d\gamma^5\psi).
\label{eca}
\end{eqnarray}
In the case of the Host action (\ref{host1}), the four-fermion interaction term is given by \cite{ar05}
\begin{eqnarray}
-\frac{3}{16}\frac{\gamma^2}{\gamma^2+1}k\beta^2\int_{\mathcal M} d^4x\det(e)(\bar\psi\gamma^d\gamma^5\psi)(\bar\psi\gamma_d\gamma^5\psi).
\label{rovelli}
\end{eqnarray}

As we can see from Eqs.~(\ref{art}) to (\ref{eca}), the bilinear term (\ref{art}) of massless
fermion fields coupled to the spin-connection (\ref{cd}) is bound to yield a non-vanishing torsion field $T^a$ (\ref{werelation1}), which is local and 
static. As a result, the spin-connection $\omega$ is no longer torsion-free and 
acquires a torsion-related spin-connection $\tilde\omega^{ab}_\mu$ (\ref{connectiont}), in addition to the torsion-free spin-connection 
$\omega^{ab}_\mu(e)$. The torsion-related spin-connection 
$\tilde\omega^{ab}_\mu$ is related to the fermion spin-current (\ref{spinc}). The quadratic term 
of the spin-connection field $\omega$ in Eq.~(\ref{rcurvature}) and the coupling between the spin-connection field 
and fermion spin-current in Eqs.~(\ref{art},\ref{cd}) lead to the quadrilinear terms of fermion fields 
in Eqs.~(\ref{eca0}) and (\ref{eca1}). Another way to see this is to treat the static torsion-related spin-connection $\tilde\omega^{ab}_\mu$ (\ref{connectiont}) as a static auxiliary field, which 
has its quadratic term and linear coupling to the spin-current of fermion fields. 
Performing the Gaussian integral of the static auxiliary field, we exactly 
obtain the quadrilinear term (\ref{eca}), in addition to the torsion-free action. 

\vskip0.5cm
\noindent
{\bf A postulation and fermion-mass generation.}
\hskip0.2cm
The gauge principle requires the action of gravitational and fermion fields be invariant under the 
diffeomorphisms of the manifold ${\mathcal M}$ and local Lorentz transformations. This leads to a pseudo-trace axial vector of 
static (non-dynamics) torsion field, which is related to the spin-current of fermion fields. The interaction between 
the pseudo-trace axial vector of torsion fields and the spin-current of fermion fields results in the four-fermion interaction 
(quadrilinear terms in massless fermion fields).
As a consequence, the gauge-invariant action consists of the torsion-free action of gravitational 
and fermion fields and four-fermion interaction. We thus postulate that it is impossible to 
have any gauge-invariant theories made by the bilinear terms of massless fermion fields coupled 
to torsion-free gravitational field, and quadrilinear terms (or high-order terms) of massless fermion fields 
must be present \cite{xue94}.      

The quadrilinear term (\ref{rovelli}) is a dimension-6 operator, and four-fermion couping is in terms of 
the Immirzi parameter $\gamma$, gravitational-coupling $k$ and gauge-coupling $\beta$. 
If quantum gravity is taken into account, we expect non-local and 
high-dimensional operators ($d>6$), which contain high-order derivatives. In this case, the torsion-related
spin-connection $\tilde\omega_\mu^{ab}$ (\ref{connectiont}) is not completely static, 
rather has a mass of the order of the Planck mass, mediating in a few 
Planck length to form effective high-dimensional operators of massless fermion fields.  
The fundamental fields $e,\omega,\psi$, and operators ${\mathcal O}(e,\omega,\psi,k,\beta)$ 
are functions of the couplings $k,\beta,\gamma$, depending on the energy-scale ${\mathcal E}$. 
First we should adopt appropriate vacuum expectational value (e.v.e) of operators 
$\langle{\mathcal O}(e,\omega,\psi,k,\beta)\rangle$ as order parameters, to describe different phases and phase transitions
in the space of couplings $k,\beta$, and parameter $\gamma$. Second, we 
try to identify the scaling-invariant 
regime (ultraviolet fix points) for the {\it low-energy limit} (${\mathcal E}/m_p\rightarrow 0$), 
where the variation of fundamental fields, couplings and operators as functions of the energy-scale 
${\mathcal E}$ is govern by renormalization group equations. Third, in such scaling-invariant 
regime we try to determine the relevant and renormalizable operators that are effective dimension-4 operators, 
to obtain an effective low-energy theory for the present Universe. 

In this Letter, we are interested 
in the one-particle-irreducible (1PI) two-point functions of fermion fields $\langle\psi(0)\bar\psi(x)\rangle$, 
since they contribute to effective operators for the energy-momentum tensor 
entering the right-hand side of the Einstein equation (\ref{eineq}) for classical gravity. 
Our goal is limited to find non-trivial fermion-mass operators $(\langle\psi\bar\psi\rangle\not\equiv 0)$ 
in terms of the four-fermion interacting strength. 
For convenience in calculations, using the Planck mass $m_p$, we rescale fermion fields $\psi\rightarrow \psi/m_p$ and 
rewrite four-fermion interaction (\ref{rovelli}) as 
\begin{equation}
g\int_{\mathcal M} d^4 x\det(e)(\bar\psi\gamma^d\gamma^5\psi)(\bar\psi\gamma_d\gamma^5\psi);
\quad g=\frac{3}{16}\frac{\gamma^2}{\gamma^2+1}k\beta^2m_p^4
\label{4fnj}
\end{equation}
where the four-fermion coupling $g$ has dimension $[m_p^2]$. We assume the gauge-coupling $\beta$ to be perturbatively small.

\vskip0.5cm
\noindent
{\bf Weak four-fermion coupling.}
\hskip0.2cm
In the weak-coupling limit $g/m_p^2 \ll 1$, the dimension-3 fermion-mass operators 
$\langle\psi\bar\psi\rangle$ identically vanish $(\langle\psi\bar\psi\rangle\equiv 0)$, 
the action (\ref{eca}) gives a weakly interacting, 
massless $SU_L(2)\otimes SU_R(2)$ fermion spectrum. We define this as the ``weak-coupling symmetric phase''.
In the intermediate range of coupling
$g$, there is a ``broken phase'' where spontaneous symmetry breaking occurs. Using large-$N_f$ expansion technique  
\footnote{$g\ll 1, N_f\gg 1$ and $gN_f$ fixed, $N_f$ is 
the number of fermion flavors} shows that the
four-fermion interaction (\ref{eca}) undergoes Nambu-Jona-Lasinio
(NJL) spontaneous chiral-symmetry breaking \cite{njl1961}. 
In this symmetry broken
phase, $SU_L(2)\otimes SU_R(2)$ chiral symmetry is violated by non-vanishing mass-operators
\begin{equation}
{1\over2}\Sigma(p)=g\int d^4x \det(e)e^{-ipx}
\langle\bar\psi(0)\cdot\psi(x)\rangle\not=0,
\label{self}
\end{equation}
where $\langle\cdot\cdot\cdot\rangle$ is the average with 
respect to the partition function $Z$ of fermionic 
part of the action (\ref{eca})
\begin{equation}
\langle\cdot\cdot\cdot\rangle={1\over Z}\int d\bar\psi d\psi
\left(\cdot\cdot\cdot\right)\exp\left\{-S_{EC}[e,\omega(e),\psi,\bar\psi]\right\}.
\label{cdot}
\end{equation}
The non-vanishing mass operator (\ref{self}) obeys the NJL gap-equation,
\begin{equation}
\Sigma(p)=\tilde g \int{d^4q\over (2\pi)^4} {\Sigma(q)\over q^2+(\Sigma(q)/m_p)^2},
\label{se}
\end{equation}
where momentum $q$ and coupling $\tilde g=gN_f/m_p^2$ are dimensionless. The critical point $\tilde g_c=8\pi^2$, which 
can be obtained by $\Sigma\rightarrow 0^+$, separates the ``broken phase'' ($\Sigma\not=0, \tilde g >\tilde g_c$) from 
the ``weak-coupling symmetric phase'' ($\Sigma\equiv 0, \tilde g <\tilde g_c$). 
$\Sigma(p)\sim {\mathcal O}(m_p)$ for $\tilde g >\tilde g_c$.
The inverse propagators of these fermions can then be written as, 
\begin{eqnarray}
S^{-1}(p)&=&i\gamma_\mu p^\mu +\Sigma(p).
\label{sb1}
\end{eqnarray}
The $SU_L(2)\otimes SU_R(2)$ chiral symmetry is realized to be $SU(2)$ 
with three Goldstone modes and a massive Higgs mode that are not
presented here. Eq.~(\ref{sb1}) corresponds to the bilinear term of massive fermion fields in the effective action, which does not preserves chiral symmetries. 

\vskip0.5cm
\noindent
{\bf Strong four-fermion coupling.}
\hskip0.3cm
We turn to the strong-coupling region, where four-fermion coupling $g$ in (\ref{eca}) is sufficiently larger
than a certain critical value $g_{\rm crit}$, bound states of three fermions (three-fermion states) 
are formed 
\begin{equation}
\Psi={m_p\over
2}(\bar\psi\cdot\psi)\psi,
\label{bound}
\end{equation}
which can be understood as a bound state of one fermion and one composite boson $(\bar\psi\cdot\psi)$. 
These three-fermion states (\ref{bound}) carry the appropriate quantum
numbers of the gauge group that accommodates $\psi$. The fermion-mass operator is $\bar\psi \Psi$ and thus massive fermion 
spectrum is consistent with the chiral symmetry $SU_L(2)\otimes SU_R(2)$.

For the purpose of understanding three-fermion states and their spectra,
we henceforth focus on the strong-coupling region $(g/m_p^2\gg 1)$. We make a rescaling of fermion fields,
\begin{equation}
\psi(x)\rightarrow g^{1/4}\psi(x),
\label{rescale}
\end{equation}
and rewrite the fermion action in terms of 
the new fermion fields
\begin{eqnarray}
S_f(x)&=&{1\over 2g^{1/2}}\left[\bar\psi(x)
\gamma_\mu \partial^\mu\psi(x)+ {\rm h.c.}\right]\label{rfa}\\
S_i(x)&=&(\bar\psi\gamma^d\gamma^5\psi)(\bar\psi\gamma_d\gamma^5\psi).\label{rs2}
\end{eqnarray}
where the gauge coupling $\beta$ is assumed to be weak. 
For the limit of strong coupling $g/m_p^2\rightarrow\infty$, the kinetic
terms $S_f(x)$ can be dropped and we calculate the partition function $Z$ (\ref{cdot}) in this strong-coupling 
limit. With $S_i(x)$ given in Eq.~(\ref{rs2}), the integral of $e^{-S_i(x)}$ is 
calculated by Grassmann anticommuting algebra,
\begin{eqnarray}
Z&=&\Pi_{x}\int
[d\bar\psi(x) d\psi(x)]\exp\left[-S_i(x)\right]
=\Pi_{x}2^{4}\not=0,
\label{stronglimit1}
\end{eqnarray}
which shows a non-trivial strong-coupling limit.  
About this strong-coupling limit (\ref{stronglimit1}), we now can perform the strong-coupling expansion of $e^{-S_f(x)}$ 
in powers of $1/g$ to calculate Green-functions of fermion fields $\langle\psi(x_1)\psi(x_2)\cdot\cdot\cdot\psi(x_n)\rangle$ \cite{ep1986,xue1997}.
In order to do integral of Grassmann anticommuting algebra, we rewrite the kinetic term $S_f(x)$ (\ref{rfa}) 
as a hopping term in the Planck spacing $a^\mu, |a^\mu|=a=1/m_p$,
\begin{eqnarray}
S_f(x)&=&{1\over 2g^{1/2}a}\left[\bar\psi(x)
\gamma_\mu \psi(x+a^\mu)-\bar\psi(x+a^\mu)
\gamma_\mu \psi(x)\right].
\label{rfaa}
\end{eqnarray}
We consider the following two-point functions that form the propagator of the composite Dirac particle  
\begin{eqnarray}
S_{LL}(x)&\equiv&\langle\psi(0),\bar\psi(x)\rangle,\label{sll}\\
S_{ML}(x)&\equiv& (2a)\langle\psi(0),\bar\Psi(x)\rangle,
\label{sml}\\
S_{MM}(x)&\equiv& (2a)^2\langle\Psi(0),\Psi(x)\rangle.
\label{smm}
\end{eqnarray}
In the lowest 
non-trivial order $O(1/g)$, we obtain the following recursion relations \cite{detail}
\begin{eqnarray}
S_{LL}(x)&=&{1\over g}\left({1\over 2a
}\right)^3\sum^\dagger_\mu S_{ML}(x+a^\mu)\gamma_\mu,\label{re1}\\
S_{ML}(x)&=&{\delta(x)\over 2g}
+{1\over g}\left({1\over 2a
}\right)\sum^\dagger_\mu S_{LL}(x+a^\mu)\gamma_\mu,
\label{re2}\\
S_{MM}(x)&=&{1\over g}\left({1\over 2a
}\right)\sum^\dagger_\mu 
\gamma_\mu\gamma_0 S^{\dagger}_{ML}(x+a^\mu)\gamma_0,
\label{re3}
\end{eqnarray}
where for an arbitrary function $f(x)$,
\begin{equation}
\sum_\mu^\dagger f(x)=\sum_\mu \left[f(x+a^\mu)-f(x-a^\mu)\right].
\nonumber
\end{equation}
Transforming these two-point functions (\ref{sll},\ref{sml},\ref{smm}) 
into momentum space, 
\begin{equation}
S_X(p)=\int d^4x e^{-ipx}S_X(x),\label{fourier}
\end{equation}
where $X=LL,ML,MM$ respectively,
we obtain three recursion relations in momentum space
\begin{eqnarray}
S_{LL}(p)&=&{1\over g}\left({i\over 4a^3
}\right)\sum_\mu \sin (p^\mu a) S_{ML}(p)\gamma_\mu,\label{rep1}\\
S_{ML}(p)&=&{1\over 2g}
+{i\over ga}\sum_\mu\sin (p^\mu a) S_{LL}(p)\gamma_\mu.
\label{rep2}\\
S_{MM}(p)&=&{1\over g}\left({i\over a
}\right)\sum_\mu \sin (p^\mu a) \gamma_\mu\gamma_0 S^{\dagger}_{ML}(p)
\gamma_0.
\label{rep3}
\end{eqnarray}
We solve these recursion relations (\ref{rep1},\ref{rep2},\ref{rep3}) and
obtain
\begin{eqnarray}
S_{LL}(p)&=&{{i\over2a}\sum_\mu\sin (p^\mu a)\gamma_\mu\over
{1\over a^2}\sum_\mu\sin^2 (p_\mu a)+M^2},\label{sll2}\\
{1\over2a}S_{ML}(p)&=&{{1\over2}M(p)\over
{1\over a^2}\sum_\mu\sin^2 (p_\mu a)+M^2},\label{slm2}\\
\left({1\over2a}\right)^2S_{MM}(p)&=&{{i\over2a}\sum_\mu\sin (p^\mu a)\gamma_\mu\over
{1\over a^2}\sum_\mu\sin^2 (p_\mu a)+M^2},\label{smm2}
\end{eqnarray}
where the chiral-invariant mass is
\begin{equation}
M=2ga.\label{fm}
\end{equation}
In addition, the two-point function,
\begin{equation}
\langle\Psi(x),\bar\psi(0)\rangle={1\over2a}\gamma_0 S_{ML}^{\dagger }(x)
\gamma_0.
\label{slm3}
\end{equation}
As a result, in the lowest non-trivial
order of the strong-coupling expansion we obtain the massive propagator of
the composite Dirac fermions, 
\begin{equation}
S(p)
={{i\over a}\sum_\mu\sin (p^\mu a)\gamma_\mu+M\over
{1\over a^2}\sum_\mu\sin^2 (p_\mu a)+M^2}\simeq {ip^\mu\gamma_\mu+M\over
p^2+M^2},
\label{sc1}
\end{equation}
for modes $p^\mu a\ll 1$. Eq.~(\ref{sc1}) corresponds to the bilinear term of massive fermion fields preserving chiral symmetries in the 
effective action, which can be written as,  
\begin{eqnarray}
S^{\rm eff}_F(e,\omega,\Psi,\bar\Psi)=\frac{1}{2}\int_{\mathcal M} d^4x
\det(e)\left[\bar\Psi e^\mu {\mathcal D}_\mu\Psi + M\bar\Psi\Psi \right]+ {\rm h.c.} ,
\label{art1}
\end{eqnarray}
and its variation with respect to the tetrat field $e_\mu^a$ gives rise to the energy-momentum tensor that contributes to 
the right-handed side of the Einstein equation (\ref{eineq}). This is the ``strong-coupling symmetric phase'', where fermion
fields are massive.


The critical value $g_{\rm crit}$ that separates the `` strong-coupling symmetric phase'' 
from the ``broken phase'' can be qualitatively determined by 
considering the complex composite scalar field,
\begin{equation}
{\cal A}=\bar\psi\cdot\psi,
\label{comps}
\end{equation}
and its propagator, i.e., the two-point function:
\begin{eqnarray}
G(x)&=&\langle{\cal A}(0),{\cal A}^{\dagger }(x)\rangle.
\label{bosonp}
\end{eqnarray}
Analogously, using the strong-coupling expansion in powers of ${1/g}$
($g/m_p^2\gg 1$), we obtain the following recursion relation in the lowest order,
\begin{equation}
G(x)={\delta(x)\over g}
+{1\over g}\left({1\over 2a
}\right)^2\sum_{\pm\mu} G(x+a^\mu).
\label{re4}
\end{equation}
Going to momentum space,
\begin{equation}
G(q)=\int d^4x e^{-iqx} G(x),
\nonumber
\end{equation}
where $q$ is the momentum of the composite scalar ${\cal A}$, we obtain 
the recursion relation (\ref{re4}) in momentum
space
\begin{equation}
G(q)={1\over g}+\left(
{1\over 2a^2}\right){1\over g}\sum_{\pm\mu}\cos (q_\mu a) G(q).
\label{rep4}
\end{equation}
As a result, we find the propagator of the massive composite scalar 
field ${\cal A}$,
\begin{eqnarray}
G(q)&=& {4\over {4\over a^2}\sum_\mu\sin^2{(q_\mu a)\over 2}
+\mu^2}\simeq {4\over q^2
+\mu^2};\label{scalar}\\
\mu^2&=& 4\left(g-{2\over a^2}\right),
\label{mas}
\end{eqnarray}
where the factor 4 is due to the four components of the composite scalar field ${\cal A}$. Thus,
$\mu^2{\cal A}{\cal A}^{\dagger}$
gives the mass term of the composite scalar field ${\cal A}$ in
the effective Lagrangian. We assume that the 1PI vertex ${\cal A} {\cal
A}^{\dagger }{\cal A}{\cal A}^{\dagger }$ is positive and the energy
of ground states of the theory is bound from the bellow. Then, we can qualitatively
discuss the second order phase transition (threshold) from the ``strong-coupling symmetric phase'' 
to the ``broken phase'' by examining the
mass term of these composite scalars $\mu^2{\cal A}{\cal A}^{\dagger}$. Spontaneous symmetry
breaking $SU(2)\rightarrow U(1)$ occurs, where $\mu^2>0$ turns to $\mu^2<0$.
Eq.~(\ref{mas}) for $\mu^2=0$ gives rise to the critical value $g_{\rm crit}$: 
\begin{equation}
g_{\rm crit}a^2=2,
\label{gc}
\end{equation}
where a phase transition takes place between
the ``strong-coupling symmetric phase'' and the ``broken phase''. 

\vskip0.5cm
\noindent
{\bf Some discussions.}
\hskip0.3cm
As already mentioned, high-dimensional operators of massless fermion fields containing high-order derivatives are expected 
if the quantum gravity is included. In this case the four-fermion 
coupling (\ref{4fnj}) and 
fermion-mass (\ref{fm}) should be functions of fermion's momentum $p^\mu$.      
Both the phase-structure and critical points for phase-transition characterized by the coupling $g$ (\ref{4fnj}) 
depend clearly also on the Immirzi parameter $\gamma$. 
Although three different phases have been differentiated, we have not been able to identify
the scaling-invariant region for the {\it low-energy limit} where some of high-dimensional operators 
receive anomalous dimensions become relevant operators of effective dimension-4, others are non-relevant and suppressed. 
We expect that the scaling-invariant regime be probably near to the critical point (\ref{gc}) so that the low-energy effective 
theory preserves chiral-gauge symmetry in high-energies and has a soft symmetry-breaking for fermion masses in low-energies. 
In this Letter we discuss the phase-structure of the Einstein-Cartan theory and a theoretical possibility to understand how 
fermion fields become massive and couple to torsion-less gravitational field. 

\comment{    
In the Einstein-Cartan theory for gravitational and massless fermion fields, static torsion fields are removed 
and quadrilinear terms in fermion fields appear, we then postulate that theories with bilinear terms (\ref{art}) 
in massless fermion fields cannot fully respect the gauge principle and absence of torsion fields, quadrilinear 
terms in fermion fields must be induced. While the dynamics of 
these quadrilinear terms, whose coupling (\ref{eca})
is related to the gravitational and gauge couplings, give rise to the effective 
bilinear terms of massive fermion fields [e.g., (\ref{art1})), which contribute to the energy-momentum tensor 
in the right-hand-side of the Einstein equation.
}

\end{document}